# Graphic-Card Cluster for Astrophysics (GraCCA) --- Performance Tests

Hsi-Yu Schive[a,] *, Chia-Hung Chien[a], Shing-Kwong Wong[a], Yu-Chih Tsai[a], Tzihong Chiueh[a]

[a] *Department of Physics, National Taiwan University, Taipei, Taiwan*
* *Corresponding Author: Tel.: +886223670358; Fax: +886223670358*
*Email: b88202011@ntu.edu.tw  (Hsi-Yu Schive)*

**Abstract**

In this paper, we describe the architecture and performance of the GraCCA system, a Graphic-Card Cluster for Astrophysics simulations. It consists of 16 nodes, with each node equipped with 2 modern graphic cards, the NVIDIA GeForce 8800 GTX. This computing cluster provides a theoretical performance of 16.2 TFLOPS. To demonstrate its performance in astrophysics computation, we have implemented a parallel direct N-body simulation program with shared time-step algorithm in this system.  Our system achieves a measured performance of 7.1 TFLOPS and a parallel efficiency of 90% for simulating a globular cluster of 1024K particles. In comparing with the GRAPE-6A cluster at RIT (Rochester Institute of Technology), the GraCCA system achieves a more than twice higher measured speed and an even higher performance-per-dollar ratio. Moreover, our system can handle up to 320M particles and can serve as a general-purpose computing cluster for a wide range of astrophysics problems.

## 1. Introduction

The gravitational N-body simulation plays a significant role in astrophysics, including planetary systems, galaxies, galactic nuclei, globular clusters, galaxy clusters, and large-scale structures of the universe. The number of particles involved (denoted as N) ranges from $O(10)$ in planetary systems to $O(10^{10})$ in cosmological simulations. Since gravity is a long range force, the main challenge of such simulation lies in the calculation of all $N^2$ pairwise interactions. Therefore anything involves particle number exceeding $10^6$ will have to employ chiefly a mean-field scheme (see below). In the case of collisional system, the evolution timescale is roughly determined by two-body relaxation time which is proportional to N/log(N) (Spitzer, 1987). It implies that the total simulation time approximately scales as $O(N^3)$ (Giersz & Heggie, 1994; Makino, 1996). Therefore, the size of such astrophysical simulation is usually limited. For example, for a CPU with 10 GFLOPS (Giga Floating Operations per Second) sustained performance, it would take more than 5 years to simulate the core collapse in a globular cluster with N = 64K.

A common way to speed up the $N^2$ force calculation is to adopt the individual time-step scheme (Aarseth, 1963) along with block time-step algorithm (McMillan, 1986; Makino, 1991). The former assigns a different and adaptive time-step to each particle. Since the characteristic time-scale in some astrophysical simulations varies greatly between a dense region and a sparse region, it is more efficient to assign an individual time-step to each particle. The latter normally quantizes the time-steps to the power of two and advances particles group-by-group. Such an algorithm is especially suitable for vector machines and cluster computers, since a group of particles may be advanced in parallel. Moreover, it also reduces the time for predicting the particle attributes.

An alternative approach to improve performance is to replace the direct-summation scheme by an approximate and efficient scheme, which has a better scaling than $O(N^2)$. Examples of such schemes include the Barnes-Hut tree code (Barnes & Hut, 1986), Particle-Mesh (PM) code (Klypin & Holtzman, 1997), Particle-Particle/Particle-Mesh ($P^3M$) code (Efstathiou & Eastwood, 1981), and Tree-Particle-Mesh (TPM) code (Xu, 1995). These schemes are efficient and can deal with a large number of particles. Accordingly, they are often used in large-scale structure simulations. The drawbacks of such schemes are the limited accuracy and the incapability to deal with close encounters, which make them inappropriate to study some physics, such as the core collapse in globular cluster.

To achieve both accuracy and efficiency, one needs a high-performance computer with direct-summation algorithm. The development of GRAPE (GRAvity piPE) (Sugimoto et al., 1990; Makino et al., 2003; Fukushige et al., 2005) is made for this purpose. It is a special-purpose hardware dedicated to the calculation of gravitational interactions. By implementing multiple force calculation pipelines to calculate multiple pairwise interactions in parallel, it achieves an ultra-high performance. The latest version, GRAPE-6, comprises 12288 pipelines and offers a theoretical performance of 63.04 TFLOPS. There is also a less powerful version, GRAPE-6A, released in 2005. It is designed for constructing a PC-GRAPE cluster system, in which each GRAPE-6A card is attached to one host computer. A single GRAPE-6A card has 24 force calculation pipelines and offers a theoretical performance of 131.3 GFLOPS. Some research institutes have constructed such PC-GRAPE clusters (Fukushige et al., 2005; Johnson & Ates, 2005; Harfst et al., 2007; MODEST[1]), where the peak performance is reported to be about 4 TFLOPS. However, the main disadvantages of such system are the relatively high cost, the low communication bandwidth, and the lack of flexibility due to its special-purpose design (Portegies Zwart et al., 2007).

By contrast, the graphic processing unit (GPU) now provides an alternative for high-performance computation (Dokken et al., 2005). The original purpose of GPU is to serve as a graphics accelerator for speeding up image processing and 3D rendering (e.g., matrix manipulation, lighting, fog effects, and texturing). Since these kinds of operations usually involve a great number of data to be processed independently, GPU is designed to work in a Single Instruction, Multiple Data (SIMD) fashion that processes multiple vertexes and fragments in parallel. Inspired by its advantages of programmability, high performance, large memory size, and relatively low cost, the use of GPU for general-purpose computation (GPGPU[2]) has become an active area of research ever since 2004 (Fan et al., 2004; Owens et al., 2005, 2007). The theoretical performance of GPU has grown from 50 GFLOPS for NV40 GPU in 2004 to more than 500 GFLOPS for G80 GPU (which is adopted in GeForce 8800 GTX graphic card) in late 2006. This high computing power mainly arises from its fully pipelined architecture plus the high memory bandwidth.

The traditional scheme in GPGPU works as follows (Pharr & Fernando, 2005; Dokken et al., 2005). First, physical attributes are stored in a randomly-accessible memory in GPU, called texture. Next, one uses the high-level shading languages, such as GLSL[3], Cg (Fernando & Kilgard, 2003), Brook (Buck et al., 2004), or HLSL[4], to program GPU for desired applications. After that, one uses graphics application programming interface (API) such as OpenGL[5] or DirectX[6] to initialize computation, to define simulation size, and to transfer data between PC and GPU memory. Note that the original design of graphic card is to render calculation results to the screen, which only supports 8-bit precision for each variable. So finally, in order to preserve the 32-bit accuracy, one needs to use a method called "frame buffer object" (FBO) to redirect the calculation result to another texture memory for further iterations. In addition, this method also makes the iterations in GPU more efficient. For example in many GPGPU applications, the entire computation may entirely reside within the GPU memory (except for initializing and storing data in hard disk), which minimizes the communication between GPU and the host computer.

---

[1] see http://modesta.science.uva.nl

[2] see http://www.gpgpu.org

[3] see http://www.opengl.org/documentation/glsl

[4] see http://msdn2.microsoft.com/en-us/library/bb509638.aspx

[5] see http://www.opengl.org

[6] see http://msdn2.microsoft.com/en-us/xna/aa937781.aspx

In February 2007, the NVIDIA Corporation releases a new computing architecture in GPU, the Compute Unified Device Architecture (CUDA) (NVIDIA, 2007), which makes the general-purpose computation in GPU even more efficient and user friendly. In comparing with the traditional graphic API, CUDA views GPU as a multithreaded coprocessor with standard C language interface. All threads that execute the same kernel in GPU are divided into several thread blocks, and each block contains the same number of threads. Threads within the same block may share their data through an on-chip parallel data cache, which is small but has much lower memory latency than the off-chip DRAMS. So, by storing common and frequently used data in this fast shared memory, it is possible to remove the memory bandwidth bottleneck for computation-intensive applications.

For hardware implementation, all stream processors in GPU are grouped into several multiprocessors. Each multiprocessor has its own shared memory space and works in a SIMD fashion. Each thread block mentioned above is executed by only one multiprocessor, so these threads may share their data through the shared memory. Take the NVIDIA GeForce 8800 GTX graphic card (NVIDIA, 2006) for example. It consists of 16 multiprocessors. Each multiprocessor is composed of 8 stream processors and has 16 KB shared memory. By allowing the dual-issue of MAD (multiplication and addition) and MUL (multiplication) instructions, this graphic card gives a theoretical computing power of 518.4 GFLOPS. Besides, it has 768 MB GDDR3 memory (named as the device memory or GPU memory) with memory bandwidth of 86.4 GB/s and supports IEEE-754 single-precision floating-point operations. By contrast, the currently most advanced memory bus, dual-channel DDR2 800, in a workstation has a memory bandwidth of 12.8 GB/s.

Scientific computations such as finite-element method and particle-particle interactions are especially suitable for GPGPU applications, since they can easily take advantage of the parallel-computation architecture of GPU. In previous works, Nyland et al. (2004) and Harris (2005) implemented the N-body simulation in GPU but with limited performance improvement. More recently, a 50-fold speedup over

Xeon CPU was achieved by using GeForce 8800 GTX graphic card and Cg shading language (Portegies Zwart et al., 2007), but it is still about an order of magnitude slower than a single GRAPE-6A card. Elsen et al. (2007) achieved nearly 100 GFLOPS sustained performance by using ATI X1900XTX graphic card and Brook shading language. Hamada and Iitaka (2007) proposed the "Chamomile" scheme by using CUDA, and achieved a performance of 256 GFLOPS for acceleration calculation only. Belleman et al. (2007) proposed the "Kirin" scheme also by using CUDA, and achieved a performance of 236 GFLOPS for acceleration, jerk, and potential calculations. Although the works of Hamada & Iitaka, and Belleman et al. have outperformed what can be achieved by a single GRAPE-6A card, these are either a sequential code that applies to a single GPU (Hamada & Iitaka, 2007) or a parallel code but only has been tested on a 2-GPU system (Belleman et al., 2007). Consequently, their performances are still incomparable to those of GRAPE-6 and GRAPE-6A cluster.

Based on these works, we have built a 32-GPU cluster named GraCCA, which is compatible to CUDA and has achieved a measured performance of about 7 TFLOPS. In this paper, we describe the architecture and performance of our GPU cluster. We first describe the hardware architecture in detail in Section 2, and then our implementation of parallel direct N-body simulation in Section 3. We discuss the performance measurements in Section 4. In Section 5, we give a theoretical performance model, and finally a discussion of comparison with GRAPE, stability of GraCCA, and some future outlook are given in Section 6.

## 2. GPU cluster

In this section, we first show the architecture of GraCCA, and then discuss the bandwidth measurement between PC and GPU memory, an issue that can be the bottleneck of scientific computation

2.1. Architecture of GraCCA

Fig. 1 shows the architecture of GraCCA, and Node 1 is enlarged for detail. The cluster consists of 16 nodes, with each node equipped with 2 graphic cards, the NVIDIA GeForce 8800 GTX, for general-purpose computation. Each graphic card has a GDDR3 memory with 768 MB and a G80 GPU. Its theoretical computing power amounts to 518.4 GFLOPS. The whole system therefore provides a theoretical computing power of 518.4 * 32 = 16.2 TFLOPS (exclusive of computation executed by CPUs). In Table 1, we list the main components of a single node in our system.

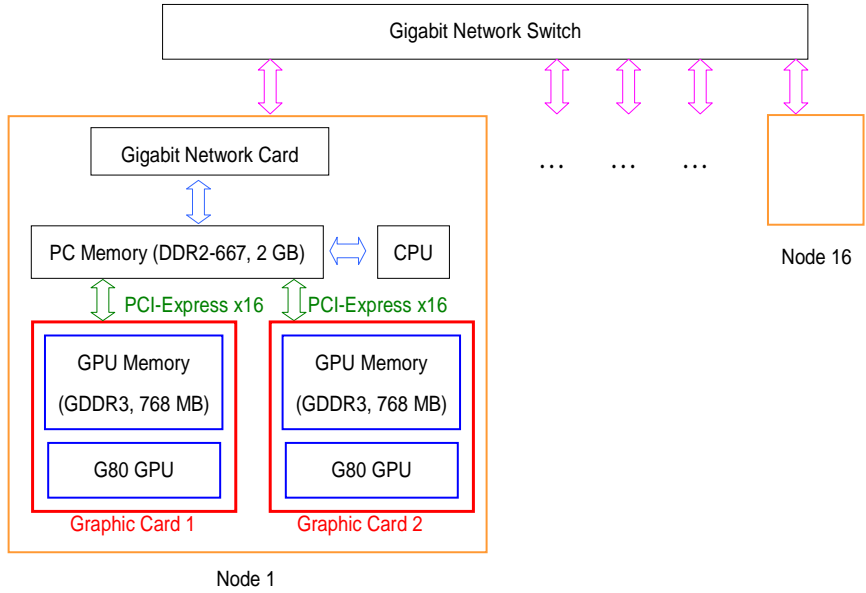

Fig. 1. The architecture of GraCCA. The node 1 is enlarged for detail. This figure is plotted by following the architecture description in Fan et al. (2004), but modified to fit the architecture of our system.

| Component | Model | Amount |
|---|---|---|
| Motherboard | Gigabyte GA-M59SLI –S5 | 1 |
| Graphic Card | Gigabyte GV-NX88X768H-RH | 2 |
| CPU | AMD Athlon 64 X2 3800 | 1 |
| Power Supply | Thermaltake Toughpower 750W | 1 |
| RAM | Transcend DDR2-667 1GB | 2 |
| Hard Disk | Seagate 80G SATAII | 1 |

Table 1

The main components of a single node in GraCCA.

Apart from graphic cards, other components are similar to those of a general PC cluster.

Each graphic card is installed in a PCI-Express x16 slot and each node is connected to a gigabit Ethernet switch. Figs. 2 and 3 are the photos of our GPU cluster and a single node, respectively. We use MPI as the API to transfer data between different CPU processes (including two processes in the same node). Each process is taken by one GPU. For transferring data between PC memory and GPU memory, we adopt CUDA library as API. Since GPU is capable of ultra-fast computation, the communication between PC and GPU memory could be a bottleneck if it is not sufficiently optimized. We illustrate this point in next section.

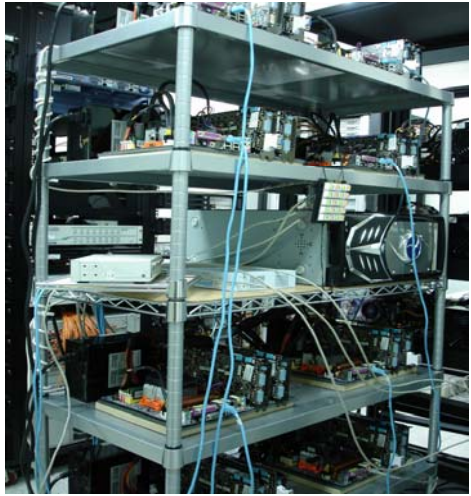 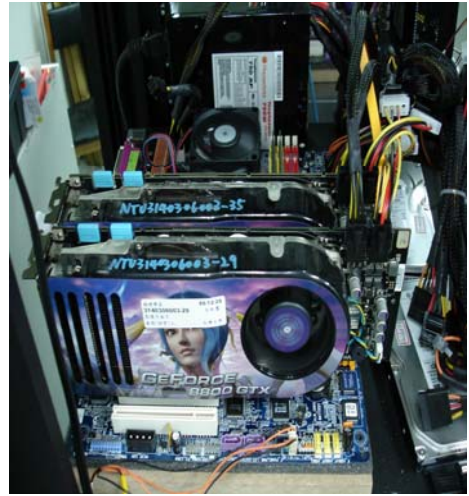

Fig. 2. The photo of GraCCA.         Fig. 3. The photo of a single node in GraCCA.

By installing two graphic cards in a single PC, we maximize the performance of a single computing node. Moreover, as shown in Section 2.2, this architecture also utilizes the total bandwidth between PC and GPU memory more efficiently.

2.2. Bandwidth between PC and GPU memory

Data transfer between PC and GPU memory contains two parts: from PC to GPU memory (downstream) and from GPU to PC memory (upstream). Although the theoretical bandwidth of PCI-Express x16 is 4GB/s in each direction, it is well known that for traditional OpenGL API, the effective bandwidth is asymmetric. So it would be more prudent to measure them separately.

Figs. 4 and 5 show the effective downstream and upstream bandwidths measured by CUDA API, as a function of the size of transferred data. The measured result is obtained by averaging over 1000 steps. In the case of single GPU, we can see that for package size > 1MB, the bandwidth in each direction both achieve about 1500 ~ 1700 MB/s. It makes no significant difference between downstream and upstream bandwidth for CUDA API. Moreover, GPUs installed in northbridge and southbridge give about the same performance (the southbridge GPU exceeds the northbridge GPU by about 150 MB/s in downstream bandwidth for large data). For a more realistic case, with multiple GPUs running parallelly, two GPUs in the same node must transfer data to and from the PC memory simultaneously. In this case, the bandwidth of each GPU reduces to about 950 MB/s for package size > 256KB, giving a total bandwidth of about 1900 MB/s. As discussed in Section 4.2, given the drop in data speed this high-speed transfer still makes the communication time between GPU and PC memory nearly negligible for a direct N-body problem.

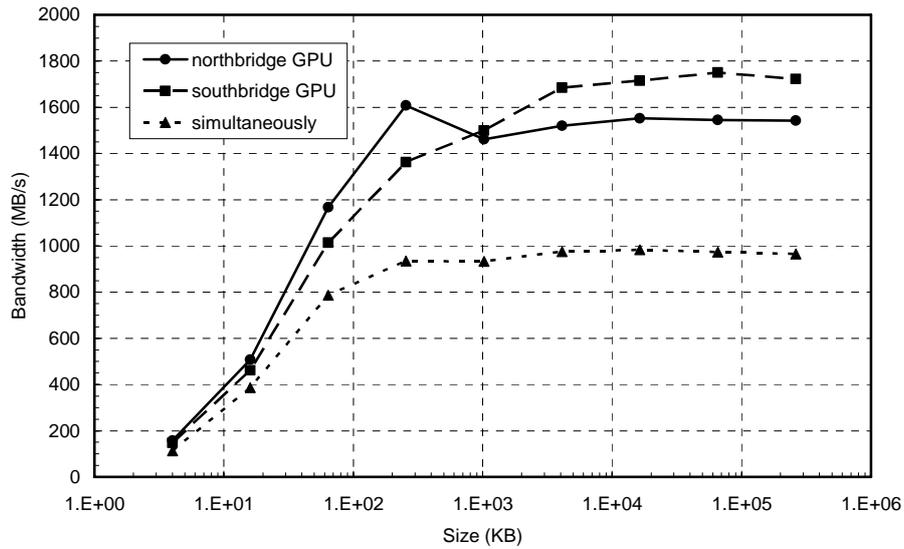

Fig. 4. The downstream bandwidths as a function of the package size. Solid, dashed, and dotted curves show the measured results of GPU in northbridge, southbridge, and when two GPUs run simultaneously, respectively.

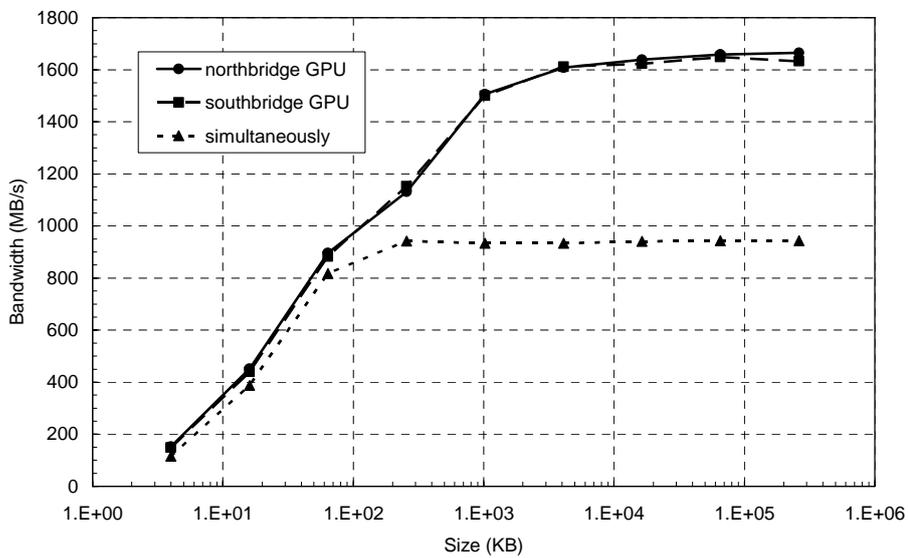

Fig. 5. The upstream bandwidths as a function of the package size. Solid, dashed, and dotted curves show the measured results of GPU in northbridge, southbridge, and when two GPUs run simultaneously, respectively.

## 3. Direct N-body simulation in GPU cluster

To demonstrate the practicability and performance of GraCCA, we have implemented the direct N-body simulation in this system. In the following, we first describe the single-GPU implementation in detail, and then follow the parallel algorithm.

3.1. Single-GPU implementation

To implement the gravitational N-body calculation in a single GPU, we follow the basic ideas of Chamomile scheme (Hamada and Iitaka, 2007) and Kirin scheme (Belleman et al., 2007), but with some modifications and a more detailed description. As described in Section 1, one of the most important features of CUDA and GeForce 8800 GTX graphic card is the small but fast on-chip shared memory. It is the key to fully explore the computing power of GPU. In addition, all threads executed in GPU are grouped into several thread blocks, and each of these blocks contains the same number of threads. For simplicity, we use the term "Grid Size (GS)" to denote the number of thread blocks, and "Block Size (BS)" to denote the number of threads within each thread block. Therefore, the total number of threads is given by GS*BS. In our current implementation, both BS and GS are free parameters which should be given before compilation. Also note that only threads within the same thread block may share their data through shared memory.

In our current implementation, only acceleration and its time derivative (jerk) are evaluated by GPU. Other parts of the program, such as advancing particles, determining time-step, and decision making, are performed in host computer. Fig. 6 shows the schematic diagram of our single-GPU implementation for acceleration and jerk calculations. Following the convention in N-body simulation, interactive particles are divided into i-particles and j-particles. The main task of GPU is to calculate the acceleration and jerk on i-particles exerted by j-particles according to the following:

$$\mathbf{a}_i = \sum_j m_j \frac{\mathbf{r}_{ij}}{(r_{ij}^2 + \varepsilon^2)^{3/2}} \qquad (1)$$

and

$$\mathbf{j}_i = \sum_j m_j \left[ \frac{\mathbf{v}_{ij}}{(r_{ij}^2 + \varepsilon^2)^{3/2}} + 3\frac{(\mathbf{v}_{ij} \cdot \mathbf{r}_{ij})\mathbf{r}_{ij}}{(r_{ij}^2 + \varepsilon^2)^{5/2}} \right], \qquad (2)$$

where m, $\mathbf{r}_{ij}$, $\mathbf{v}_{ij}$, $\mathbf{a}$, $\mathbf{j}$, and $\varepsilon$ are mass, relative position, relative velocity, acceleration, jerk, and softening parameter. To make it more clearly, we use $P_{ks}$ to denote the pairwise interaction between $s^{th}$ i-particle and $k^{th}$ j-particle. So, to match the CUDA programming model and extract the maximum performance of GPU, all $N^2$ pairwise interactions are grouped into $(N/BS)^2$ groups (denoted as $G_{mn}$, m = 1, 2, ..., N/BS, n = 1, 2, ..., N/BS). Each group $G_{mn}$ contains $BS^2$ pairwise interactions between i-particles and j-particles. It may be expressed as

$$G_{mn} = \{P_{ks} \mid (m-1)\cdot BS < k \leq m\cdot BS; (n-1)\cdot BS < s \leq n\cdot BS\}. \qquad (3)$$

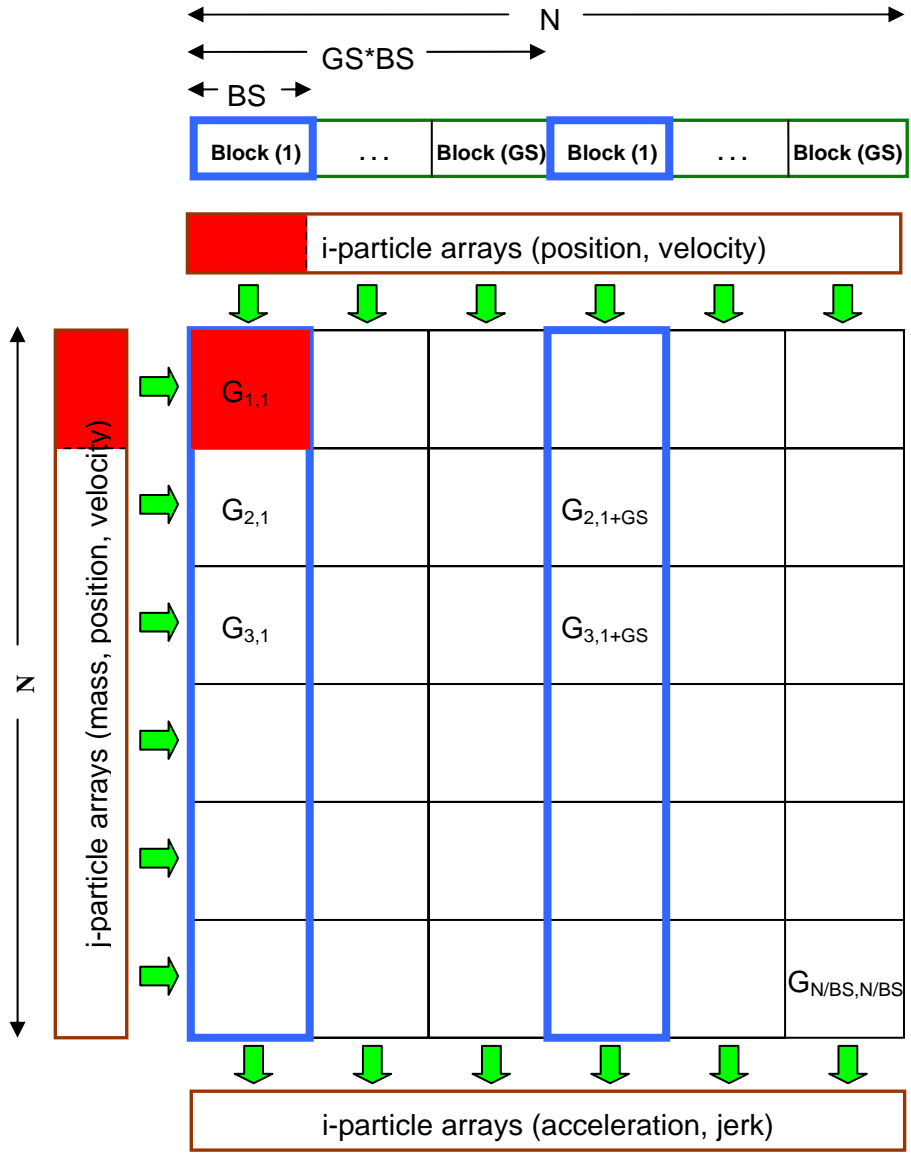

Fig. 6. The schematic diagram of our single-GPU implementation for acceleration and jerk calculations. The interaction groups computed by Block(1) are highlighted with blue border. The red regions in i-particle and j-particle arrays are the particles used to compute the group $G_{1,1}$.

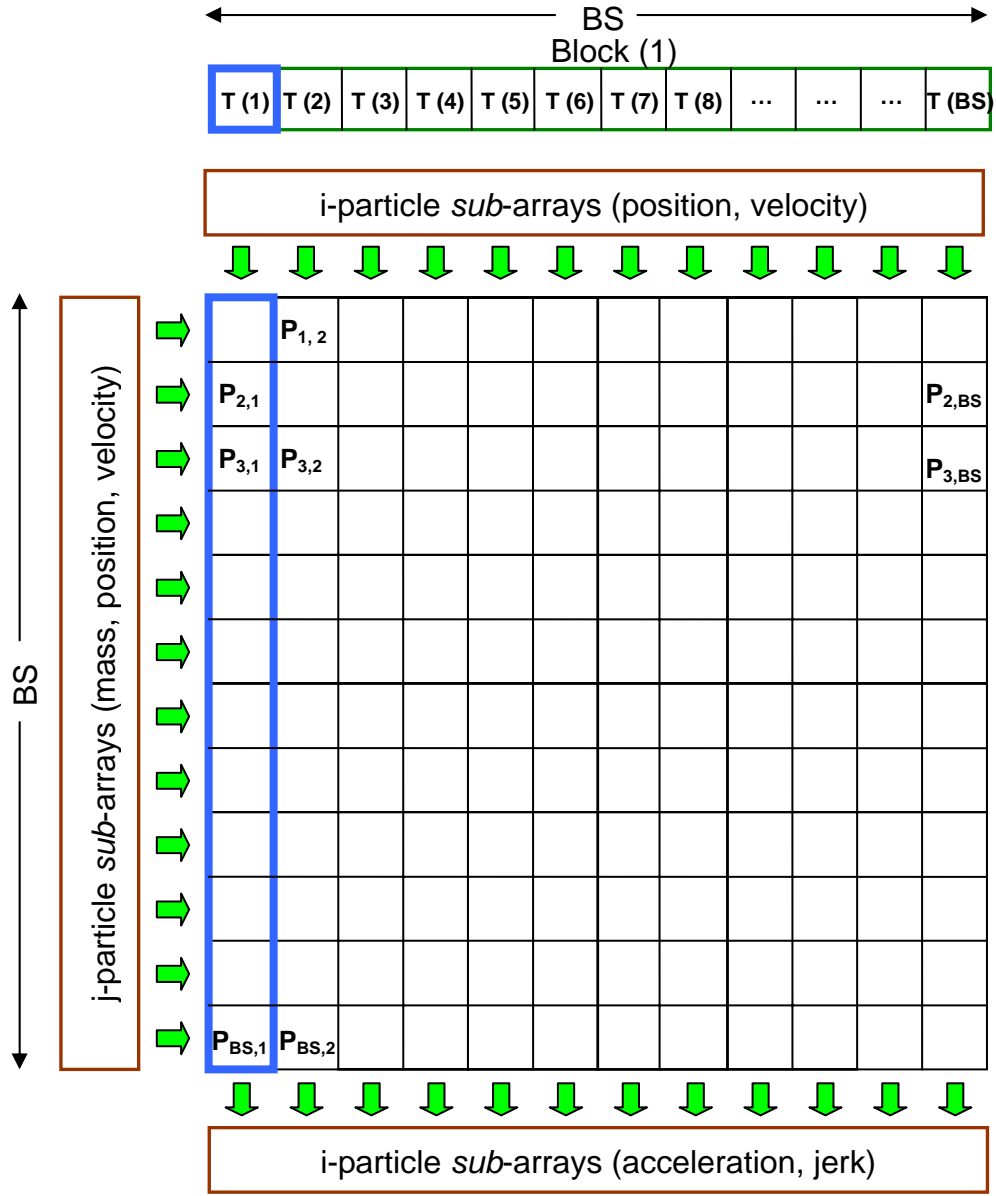

Fig. 7. The schematic diagram of the evaluation of group $G_{1,1}$ in Fig. 6. The T(s) stands for the $s^{th}$ thread within the Block(1). The interaction pairs computed by T(1) are highlighted with blue border.

Groups within the same column are computed by the same thread block sequentially. In other words, for those i-particles belong to the same column of groups, the acceleration and jerk are evaluated group-by-group by a single thread block. For the case that N/BS > GS, a thread block should evaluate more than one column of $G_{ij}$. For example, $G_{1,1}$, $G_{2,1}$, …, $G_{N/BS,1}$, $G_{1,1+GS}$, $G_{2,1+GS}$, …, $G_{N/BS,1+GS}$, …, are evaluated by Block(1); $G_{1,2}$, $G_{2,2}$, …, $G_{N/BS,2}$, $G_{1,2+GS}$, $G_{2,2+GS}$, …, $G_{N/BS,2+GS}$, …, are evaluated by Block(2), etc.

The evaluation of group $G_{1,1}$ in Fig. 6 is shown in detail in Fig. 7. Block(1) is comprised of BS threads, and each Thread(s) evaluates the acceleration and jerk on $s^{th}$ i-particle exerted by j-particles. For example, $P_{2,1}$, $P_{3,1}$, ..., $P_{BS,1}$ are evaluated by Thread(1); $P_{1,2}$, $P_{3,2}$, ..., $P_{BS,2}$ are evaluated by Thread(2), etc. This kind of computation decomposition fully exploits the immense parallel computing power of modern GPU. Besides, since threads within the same thread block may share their data through fast shared memory, each thread only needs to load one j-particle (the $s^{th}$ j-particle) into the shared memory. It reduces the number of data transfers between device memory and shared memory, which has much higher memory latency and lower bandwidth than the on-chip memory. Moreover, since the number of pairwise force calculations in $G_{mn}$ is proportional to $BS^2$, but the number of data loading from device memory to shared memory is proportional to BS, we could further eliminate this memory bandwidth bottleneck by having larger BS (128 for example). On the other hand, because different thread evaluates force on different i-particle, we may store the information of i-particles in per-thread registers instead of shard memory.

The calculation procedure of a force loop may be summarized as following:

(1) The host computer copies the data of i-particles and j-particles from PC memory to device memory through PCI-Express x16 slot.
(2) Each thread loads the data of i-particle into registers based on one-to-one correspondence.
(3) Each thread loads the data of j-particle into shared memory based on one-to-one correspondence.
(4) Each thread block evaluates one group of pairwise interactions ($G_{mn}$).
(5) Repeat (3)-(4) for m = 1, 2, ..., N/BS.
(6) Repeat (2)-(5) R times if (R-1)*GS < N/BS $\leqq$ R*GS.
(7) GPU copies the acceleration and jerk on i-particles from device memory back to PC memory through PCI-Express x16 slot.

Note that by iterating over j-particles first, all data of i-particles may stay in the same registers during the calculation of whole column of $G_{mn}$. Moreover, when switching from one $G_{mn}$ to another, each thread only needs to reload 7 variables (mass, position, and velocity of j-particles) instead of 12 (position, velocity, acceleration, and jerk of i-particles) from the device memory. So, it reduces the communication time between device memory and on-chip memory and results in a better performance, especially for small number of particles.

Finally, to integrate the orbits of particles, currently we adopt the fourth-order Hermite scheme (Makino & Aarseth, 1992) with shared time-step algorithm. For time-step determination, we first use the formula (Aarseth, 1985)

$$dt = \left( \nu \frac{|\mathbf{a}||\ddot{\mathbf{a}}| + \mathbf{j}^2}{|\mathbf{j}||\dddot{\mathbf{j}}| + \ddot{\mathbf{a}}^2} \right)^{1/2}, \tag{4}$$

where $\nu$ is an accuracy parameter, to evaluate the time-step for each particle, and then adopt the minimum of them as the shared time-step.

3.2. Parallel algorithm

To parallelize the direct N-body simulation in GraCCA, we adopt the so called "Ring Scheme". In this scheme, all GPUs are conceptually aligned in a circle. Each GPU contains a subset of $N/N_{gpu}$ i-particles (denoted as Sub-I, $N_{gpu}$ denotes the total number of GPUs). Besides, j-particles are also divided into $N_{gpu}$ subsets (denoted as Sub-J), and a force loop is composed of $N_{gpu}$ steps. During each step, each GPU evaluates the force from a Sub-J on its own Sub-I, and then transfer the data of Sub-J between different GPUs.

The calculation procedure of a force loop may be summarized as following:

(1) Initialize the acceleration and jerk backup arrays of each Sub-I as zeros.
(2) Copy the mass, position, and velocity arrays of each Sub-I to that of Sub-J.
(3) Use GPU to compute the acceleration and jerk on Sub-I exerted by the current Sub-J.
(4) Use CPU to sum the computing results of GPU with the backup arrays.
(5) Send the data of Sub-J to the GPU in clockwise direction and receive the data of Sub-J from the GPU in counterclockwise direction. Replace the data of current Sub-J by the received data.
(6) Repeat (3)-(5) $N_{gpu}$ times.

Note that in this scheme, we may use the non-blocking send and receive (ISEND and IRECV in MPI) to start the data transfer before step (4). The next force loop will wait until the data transfer is complete. By doing so, we could reduce the network communication time since it would be partially overlapped with the force computation (Dorband et al., 2003).

## 4. Performance

In this section, we discuss the performance of GraCCA for direct N-body simulation. For all performance-testing simulations, we used the Plummer model with equal-mass particles as initial condition and adopted the standard units (Heggie &

Mathieu, 1986), where gravitational constant G is 1, total mass M is 1, and total energy E is -1/4. This initial condition is constructed by using the software released by Barnes (1994). For software, we used Linux SMP kernel version 2.6.16.21-0.8, gcc version 4.1.0, CUDA Toolkit version 0.8 for Linux x86 32-bit, CUDA SDK version 0.8.1 for Linux x86 32-bit, and Linux Display Driver version 97.51.

In the following, we first discuss the optimization of GS and BS. We then assess the performance of single-GPU system, and finally the performance of multi-GPU system.

4.1. Optimization of GS and BS

As mentioned in Section 3.1, both GS (number of thread blocks) and BS (number of threads within each thread block) are free parameters in our current implementation. In theory, in order to maximize the utilization of GPU resources, both GS and BS should be chosen as large as possible. But on the other hand, a larger BS would introduce a higher cumulative error (Hamada and Iitaka, 2007). So, it would be necessary to determine the optimized values of GS and BS.

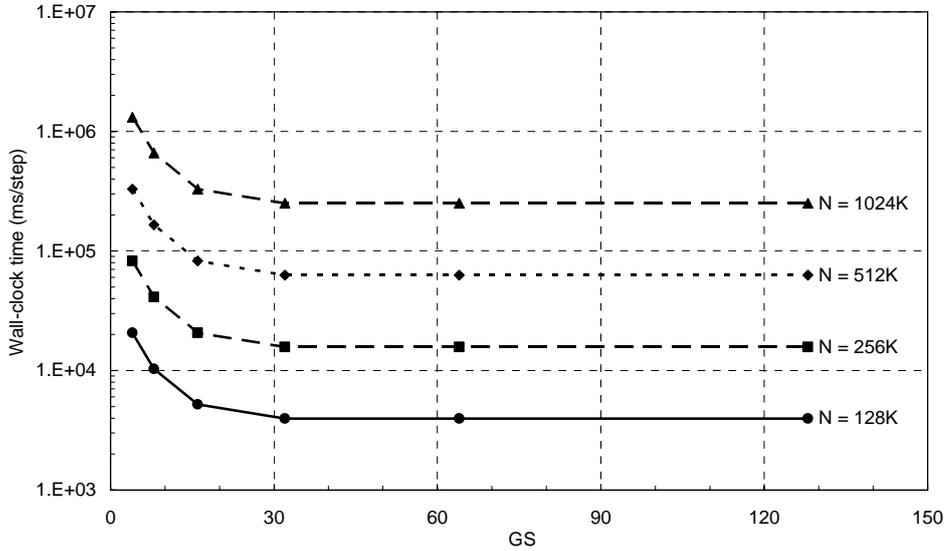

Fig. 8. The wall-clock time per step as a function of the Grid Size (GS). Solid, dashed (square), dotted, and dashed (triangle) curves show the measured results for N = 128K, 256K, 512K, and 1024K, respectively.

Fig. 8 shows the calculation time per step as a function of GS for different number of particles. BS is set to 128. It can be seen that for GS $\leqq$ 16, the calculation time per step is inversely proportional to GS. This result is consistent with the architecture of

GeForce 8800 GTX, which has exactly 16 multiprocessors. Since each thread block is executed by only one multiprocessor, executing a kernel in GPU with GS $\leqq$ 16 will result in 16-GS "idle" multiprocessors. On the other hand, for GS = n*16, n = 2, 3, 4, ..., each multiprocessor processes more than one thread block concurrently. It enables a more efficient utilization of GPU resources. As shown in Fig. 8, a single-GPU system is able to achieve its maximum performance for GS $\geqq$ 32.

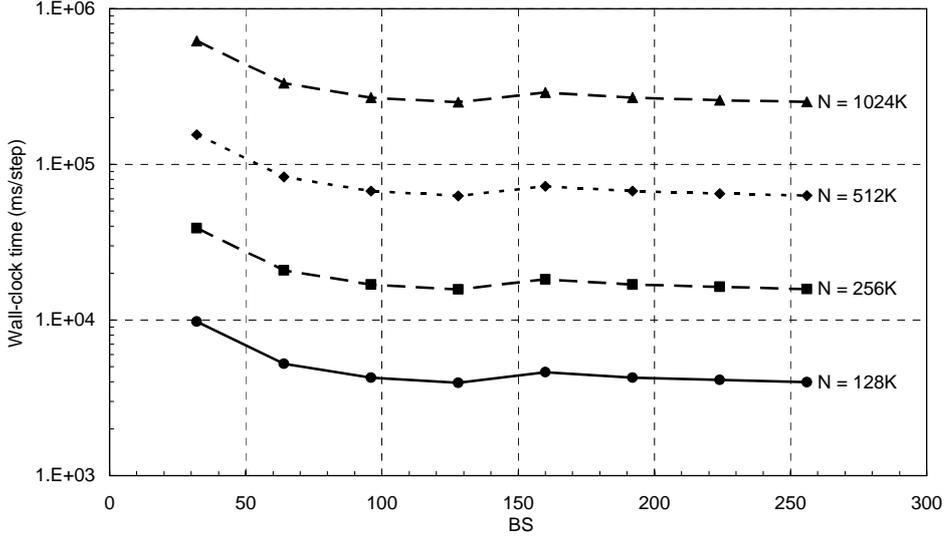

Fig. 9. The wall-clock time per step as a function of the Block Size (BS). Solid, dashed (square), dotted, and dashed (triangle) curves show the measured results for N = 128K, 256K, 512K, and 1024K, respectively.

Fig. 9 shows the calculation time per step as a function of BS for different number of particles. GS is set to 32. It can be seen that for BS $\geq$ 96 (except for BS = 160), it approaches the maximum performance. The best performance occurs for BS = 128 in our current implementation. Note that for BS = 160, the performance drops about 13 percent. It is mainly due to the current implementation of CUDA and graphic driver, and may be improved in future versions.

Accordingly, we adopt (GS, BS) = (32,128) for all performance tests in the following sections (except for the cases when $N/N_{gpu} < 4K$).

4.2. Single-GPU performance

Fig. 10 shows the calculation time per step versus the total number of particles. The measured result is obtained by averaging over 1000 steps for N < 13K and over 100 steps for N $\geqq$ 13K. Calculation time for initialization procedure is excluded. For the GRAPE system (Makino et al., 2003; Fukushige et al., 2005), a time estimate is provided to evaluate the system speed.  In a similar fashion, the total calculation time per step for a single-GPU system can be expressed as

$$T_{single} = T_{host} + T_{PCIe} + T_{GPU},  \qquad(5)$$

where $T_{host}$ is the time for host computer to predict and correct particles, as well as to determine the next time-step, $T_{PCIe}$ is the time for transferring data between PC and GPU memory through PCI-Express x16 slot, and $T_{GPU}$ is the time for GPU to calculate the acceleration and jerk.

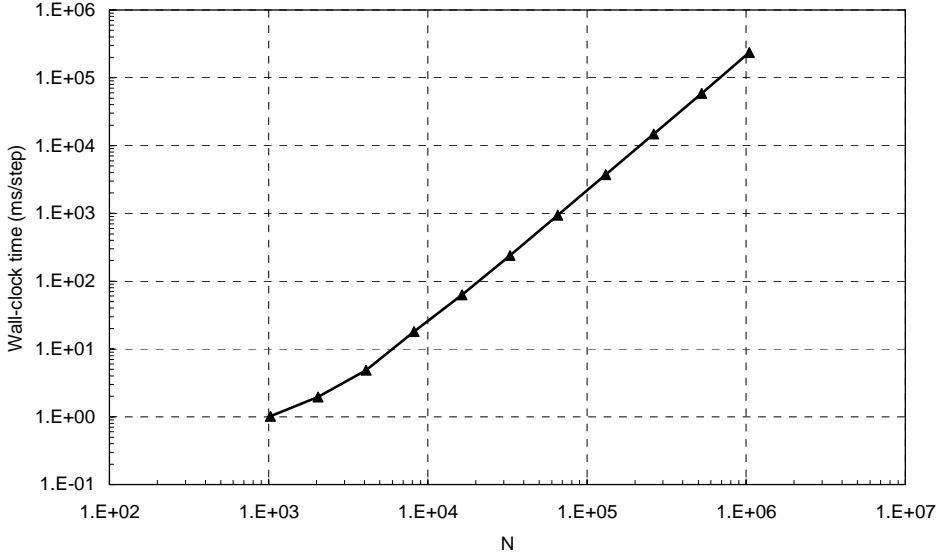

Fig. 10. The wall-clock time per step as a function of the number of particles (N) in the single-GPU system.

It is clear from Fig. 10 that for $N \geqq 4K$, the performance curve has a slope of 2, which is the signature of $N^2$ calculation. It also verifies that for a large number of N, both $T_{host}$ and $T_{PCIe}$ are negligible. But for $N < 4K$, insufficient number of particles results in inefficient utilization of GPU resources. Moreover, the time for communication between PC and GPU memory and for computation in host computer become non-negligible. These factors further reduce the performance. We will describe the performance modeling for single-GPU calculation in detail in Section 5.1.

Fig. 11 gives the measured performance in GFLOPS as a function of the total number of particles. The performance in GFLOPS is defined as

$$P_{GFLOPS} = \frac{N^2 N_{FLOP}}{T_{single}} \cdot 1024^{-3},  \qquad(6)$$

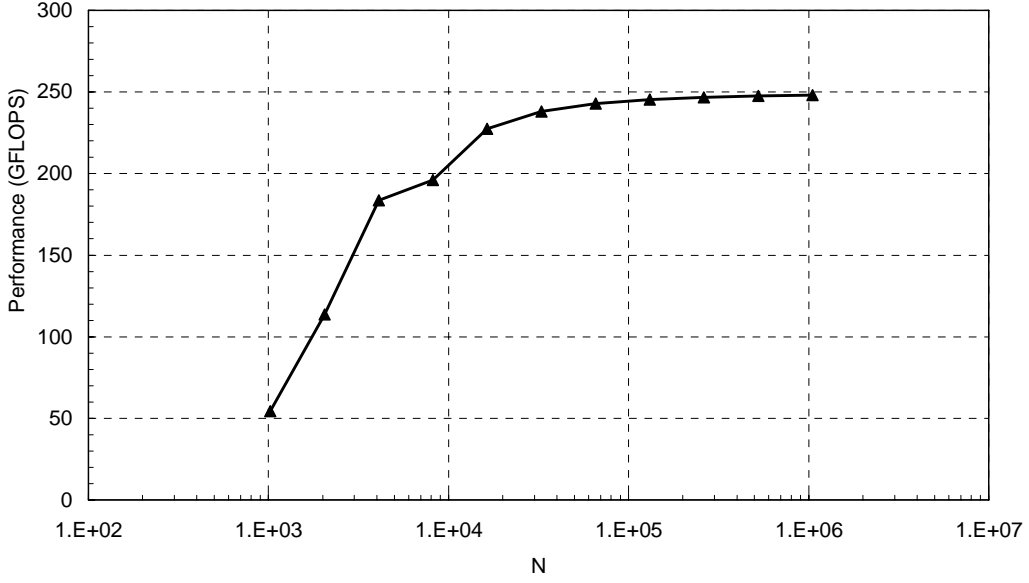

Fig. 11. The measured performance in GFLOPS as a function of the number of particles (N) in the single-GPU system.

where $N_{FLOP}$ is the total number of floating-point operations for one pairwise acceleration and jerk calculation, and $T_{single}$ is the average calculation time per step in single-GPU system. Here we adopt $N_{FLOP} = 57$ (Makino et al., 2003; Fukushige et al., 2005) in order to compare to the result of the GRAPE system. As discussed above, the performance drops for small values of N (N < 4K) due to data communication, host computer computation, and insufficient threads in GPU. On the other hand for $N \geqq$ 16K, the single-GPU system approaches its peak performance, which is about 250 GFLOPS for acceleration and jerk calculations. We note that the performance of the single-GPU system is limited by the computing power of GPU itself. Also note that here we use $T_{single}$ instead of $T_{GPU}$ in Eq. (6) for calculating GFLOPS. It makes Fig. 11 more practical and illustrative since $T_{single}$ and $T_{GPU}$ could be significantly different for small N (see Fig. 18).

4.3. Multi-GPU performance

Fig. 12 shows the calculation time per step versus the total number of particles for different number of GPUs ($N_{gpu}$). Six curves denote the 1-, 2-, 4-, 8-, 16-, and 32-GPU systems, respectively. For the multi-GPU system, the total calculation time per step may be expressed as

$$T_{multi} = T_{host} + T_{PCIe} + T_{GPU} + T_{net},  \qquad (7)$$

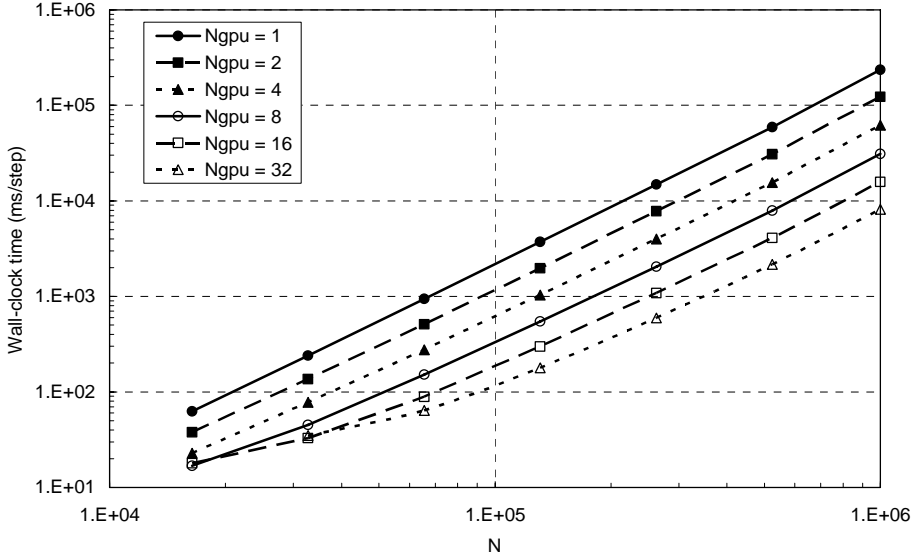

Fig. 12. The wall-clock time per step as a function of the total number of particles (N) for different number of GPUs ($N_{gpu}$). The cases of $N_{gpu}$ = 1, 2, 4, 8, 16, and 32 are plotted, respectively.

where $T_{host}$, $T_{PCIe}$, $T_{GPU}$ are defined in the same way as Section 4.2, and $T_{net}$ is the time for transferring data between different nodes through the gigabit network. Note that for a dual-GPU system, the result is measured by two GPUs installed in the same node, which provides higher communication bandwidth between the two GPUs.

In Fig. 12, all six curves have slope of 2 for $N/N_{gpu} \geqq 4K$, which is consistent with Fig. 10. It shows that for large numbers of N, $T_{host}$, $T_{PCIe}$, and $T_{net}$ are all negligible, giving $T_{multi} \sim T_{GPU}$. We will describe the performance modeling for multi-GPU calculation in detail in Section 5.2.

Fig. 13 shows the results of performance measurements in GFLOPS as a function of the total number of particles for different numbers of GPUs. We can see that for $N/N_{gpu} \geqq 16K$, each system with different number of GPUs approaches their peak performance. Moreover, it demonstrates a great scalability of our system. The maximum performance of the multi-GPU system is still limited by the computing power of GPU itself. For the 32-GPU case, the system achieves a total computing power of 7.151 TFLOPS, in which case each GPU achieves a performance of 223 GFLOPS. It is about 89 percent of the peak performance in a single-GPU system.

In Figs. 12 and 13, the crossover points indicate that the system with more GPUs becomes marginally faster or even slower than the system with fewer GPUs. All these points appear when $N/N_{gpu} = 1K$. This result is consistent with Fig. 11, since when the number of particles changes from 2K to 1K, the performance of single GPU drops more than 50%.

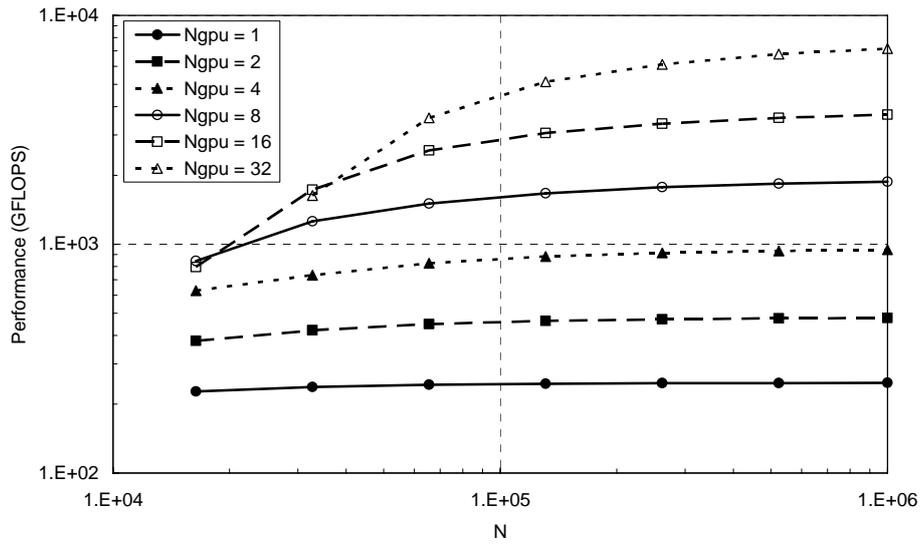

Fig. 13. The measured performance in GFLOPS as a function of the total number of particles (N) for different number of GPUs ($N_{gpu}$). The cases of $N_{gpu}$ = 1, 2, 4, 8, 16, and 32 are plotted, respectively.

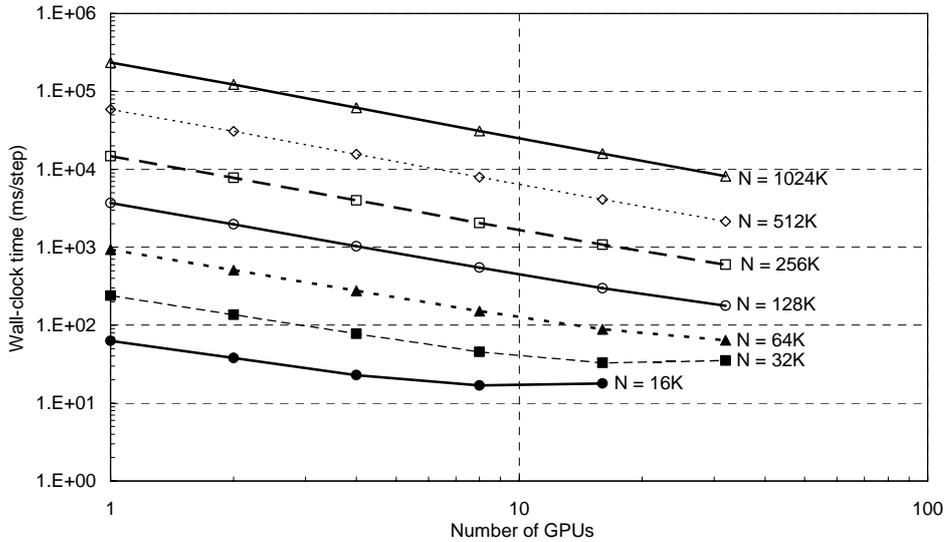

Fig. 14. The wall-clock time per step versus the number of GPUs for different number of particles (N). The cases of N = 16K, 32K, 64K, 128K, 256K, 512K, and 1024K are plotted, respectively.

Fig. 14 shows the calculation time per step as a function of the number of GPUs for different N. For $N/N_{gpu} \geqq 4K$, all curves have slope of about -1, which indicates that the calculation time per step is roughly inversely proportional to the number of GPUs. It further verifies the great scalability of our GPU cluster, especially for $N \geqq 128K$. For the N = 1024K case, it has a slope of -0.974. The calculation speed of 32-GPU system is 28.8 times faster than single-GPU system, in which the parallel efficiency achieves 90.1%. Fig. 15 shows the speedup factor versus the number of GPUs for different N, where the speedup factor s is defined as

$$s(N, N_{gpu}) = \frac{T_{single}(N)}{T_{multi}(N, N_{gpu})}. \tag{8}$$

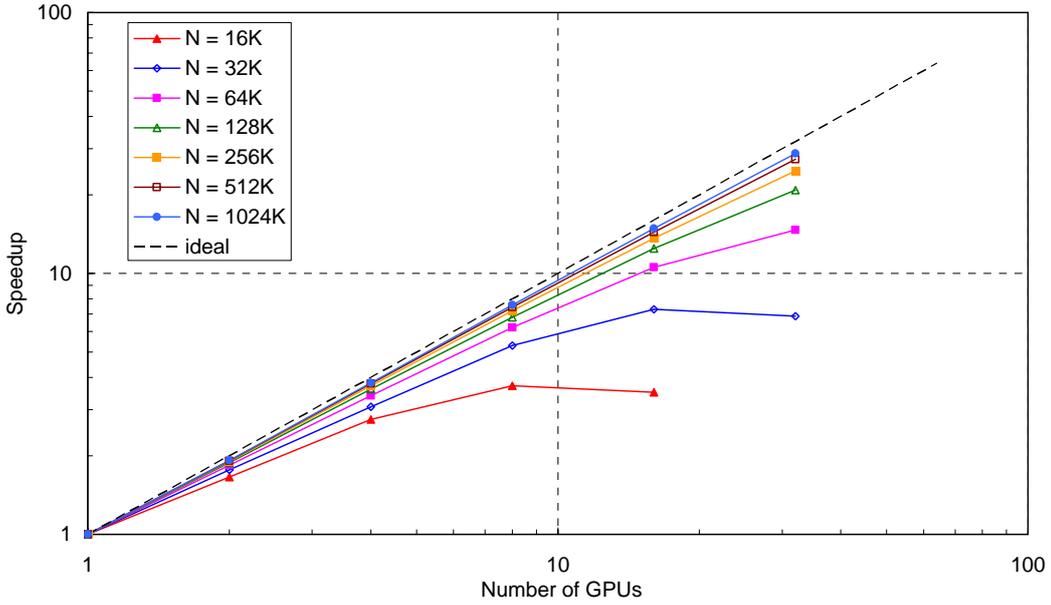

Fig. 15. The speedup factor versus the number of GPUs for different number of particles (N). The cases of N = 16K, 32K, 64K, 128K, 256K, 512K, and 1024K are plotted, respectively. The block dashed curve indicates the case of ideal speedup.

To demonstrate the GPU's ability for conducting the most time-consuming astrophysical computation, in Fig. 16 we show the temporal evolution of core density in the Plummer model up to N = 64K. The core density is estimated by using the method proposed by Casertano and Hut (1985), and with a faster convergence property suggested by McMillan et al. (1990). The time (x axis) in Fig. 16 is scaled by 212.75*log(0.11N)/N (Giersz & Heggie, 1994). Note that to capture the post-collapse behavior (e.g., gravothermal oscillation), one needs an alternative integration scheme such as KS regularization (Mikkola & Aarseth, 1998) to handle close two-body encounters and stable binaries (Makino, 1996). Currently this scheme is not yet implemented in our program.

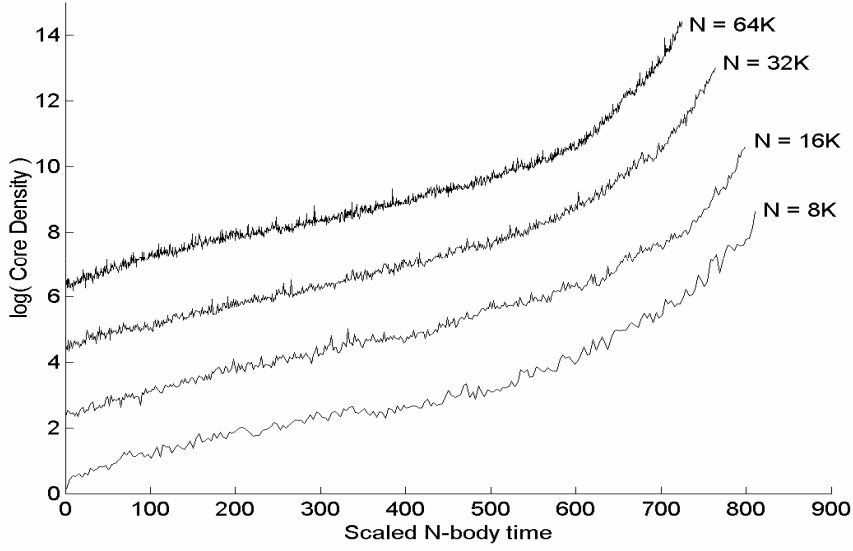

Fig. 16. The temporal evolution of core density in the Plummer model. The cases of N = 8K, 16K, 32K, and 64K are plotted and vertically shifted by 0, 2, 4, and 6 units, respectively.

## 5. Performance modeling

In this section, we construct a performance model of direct N-body simulation in GraCCA, and compare that to the measured performance. The performance model we adopt is similar to that of GRAPE-6 (Makino et al., 2003) and GRAPE-6A (Fukushige et al., 2005), but modified to fit the architecture of our system. In the following, we first present a performance model of the single-GPU system, and then follow the model of cluster system.

5.1. Performance modeling of single-GPU system

As mentioned in Section 4.2, the calculation time per step for direct N-body simulation in a single-GPU system ($T_{single}$) may be modeled by Eq. (5) : $T_{single}(N) = T_{host}(N) + T_{PCIe, single}(N) + T_{GPU}(N)$. $T_{host}$ is the time spent on the host computer. In our current implementation, it may be written as

$$T_{host}(N) = T_{pred}(N) + T_{corr}(N) + T_{timestep}(N) , \qquad (9)$$

where $T_{pred}$ is the time for prediction, $T_{corr}$ is the time for correction, and $T_{timestep}$ is the time for time-step determination. All of these operations are roughly proportional to the number of particles, so we may rewrite Eq. (9) as

$$T_{host}(N) = (t_{pred} + t_{corr} + t_{timestep}) \cdot N \equiv t_{host} \cdot N , \qquad (10)$$

where the lower-case letter "t" represents the computation time "per particle". This number is mainly determined by the computing power of host computer, and is roughly the same for different N. So in our performance model, we take $t_{host}$ as a constant and $T_{host}(N)$ is directly proportional to N. $T_{PCIe, single}$ is the time spent on data transfer in PCI-Express x16 lanes. Since the effective bandwidth between PC and GPU memory in a single-GPU case is different from the multi-GPU case (see Section 2.2), here we use the subscript "single" to emphasize the difference. $T_{PCIe, single}$ may be written as

$$T_{PCIe,single}(N) = T_i(N) + T_j(N) + T_{force}(N) = (t_i + t_j + t_{force}) \cdot N \\ \equiv t_{PCIe,single} \cdot N, \quad (11)$$

where $T_i$ is the time for transferring i-particle position and velocity downstream to GPU memory, $T_j$ is the time for transferring j-particle mass, position, and velocity downstream to GPU memory, and $T_{force}$ is the time for transferring i-particle acceleration and jerk upstream to PC memory. The lower-case "t" represents the communication time per particle. They could be written as $t_i = 24/BW_{down}$, $t_j = 28/BW_{down}$, and $t_{force} = 24/BW_{up}$, where $BW_{down}$ and $BW_{up}$ represent the downstream and upstream bandwidth, respectively. So, by measuring $BW_{down}$ and $BW_{up}$ (see Fig. 4 and 5), we may estimate $t_{PCIe, single}$. Finally, $T_{GPU}$ is the time spent on force calculation in GPU. It may be expressed as

$$T_{GPU} = t_{pair} \cdot N^2, \quad (12)$$

where $t_{pair}$ is the calculation time for a single pairwise interaction. Note that $T_{GPU}$ scales with $N^2$. The measured results of $t_{host}$, $t_{PCIe, single}$, and $t_{pair}$ are recorded in unit of millisecond in Table 2.

| $t_{host}$ | $t_{PCIe, single}$ | $t_{PCIe, multi}$ | $t_{net}$ | $t_{pair}$ |
|---|---|---|---|---|
| $2.746 * 10^{-4}$ | $4.745 * 10^{-5}$ | $7.606 * 10^{-5}$ | $2.725 * 10^{-4}$ | $2.139 * 10^{-7}$ |

Table 2

Measured results of performance parameters in single- and multi-GPU systems (in unit of millisecond).

Fig. 17 shows the wall-clock time per step predicted by this performance model (denoted as model 1), along with the measured performance for comparison. The agreement between model 1 and the measured result is quite good, except for the case with small number of particles (N < 4K). The discrepancy is originated from the lower

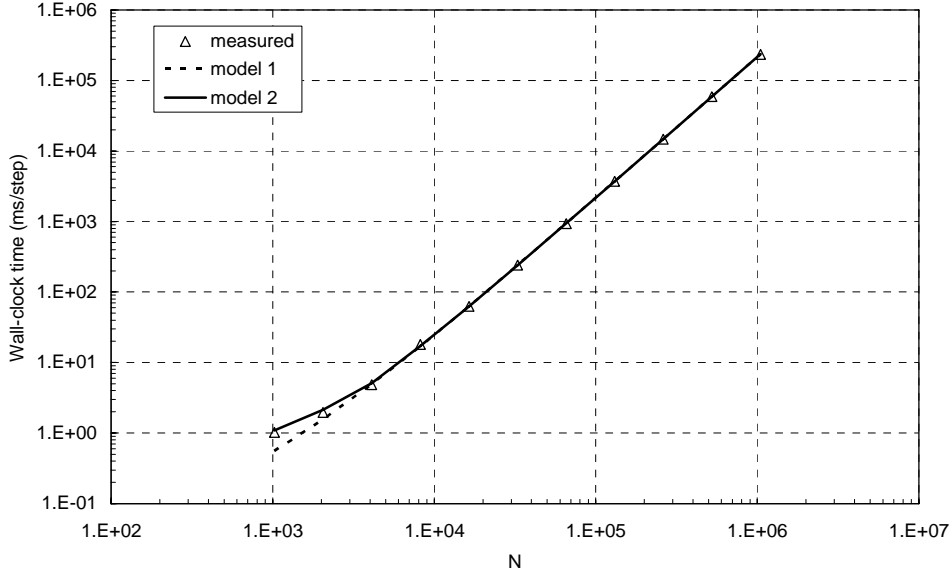

Fig. 17. The wall-clock time per step as a function of the number of particles (N) in the single-GPU system. Dotted and solid curves denote the results predicted by model 1 and model 2, respectively. Open triangles are the measured results.

PCI-Express x16 bandwidth and the less efficient utilization of GPU resource for small N (see Fig. 4, 5, and 11). Since the numbers recorded in Table 2 are the optimum values, it would be more prudent to define efficiency factors both for $t_{PCIe,single}$ and $t_{pair}$ for small N (denoted as model 2). In practice, we may rewrite Eq. (5) as

$$T_{single}(N) = t_{host} \cdot N + \frac{t_{PCIe,single} \cdot N}{f_{PCIe,single}(N)} + \frac{t_{pair} \cdot N^2}{f_{pair}(N)}, \qquad (13)$$

where

$$f_{PCIe,single}(N) = \begin{cases} 0.207\log(N) - 1.152 & (N \leq 16K) \\ 1 & (N > 16K) \end{cases},$$

$$f_{pair}(N) = \begin{cases} 0.454\log(N) - 2.793 & (N \leq 4K) \\ 1 & (N > 4K) \end{cases}.$$

$f_{PCIe,single}(N)$ and $f_{pair}(N)$ are efficiency factors for $t_{PCIe,single}$ and $t_{pair}$, respectively. These factors are purely empirical and determined by fitting to the measured results of $t_{PCIe,single}$ and $t_{pair}$. Note that although $f_{PCIe,single}$ and $f_{pair}$ are discontinuous at N = 16K and 4K, respectively, they are negligible since these discontinuities have influences on predicted performance by less than 2%. The wall-clock time per step predicted by model 2 is also presented in Fig. 17. It is clear that for N < 4K, model 2 is in better agreement with the measured performance than model 1.

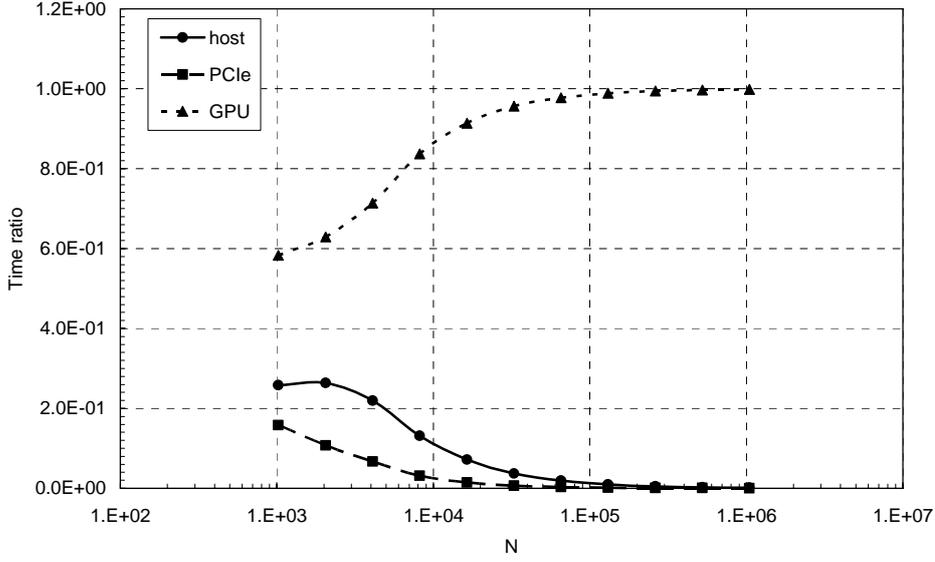

Fig. 18. The relative ratios of $T_{host}$, $T_{PCIe, single}$, and $T_{GPU}$ as a function of the number of particles (N) predicted by model 2. Solid, dashed, and dotted curves denote the $T_{host}/T_{single}$, $T_{PCIe, single}/T_{single}$, and $T_{GPU}/T_{single}$, respectively.

Fig. 18 shows the relative ratios of $T_{host}$, $T_{PCIe,single}$, and $T_{GPU}$ in model 2. Since $T_{GPU}$ scales with $N^2$, but $T_{host}$ and $T_{PCIe,single}$ scale with N, $T_{GPU}/T_{single}$ would increase with N. This feature is clearly verified in Fig. 18. $T_{GPU}/T_{single}$ reaches 71% for N = 4K and 91% for N = 16K. These predicted ratios are consistent with the timing measurement discussed in Section 4.2.

5.2. Performance modeling of multi-GPU system

Following the performance model of the single-GPU system, the calculation time per step in our GPU cluster may be modeled as

$$T_{multi}(N,n) = t_{host} \cdot n + \frac{t_{PCIe,multi} \cdot N}{f_{PCIe,multi}(n)} + \frac{t_{pair} \cdot N \cdot n}{f_{pair}(n)} + T_{net}(N), \qquad (14)$$

where $n \equiv N/N_{gpu}$ is number of i-particles held by each GPU. The subscripts "multi" in $t_{PCIe,multi}$ and $f_{PCIe,multi}$ are used to highlight the difference of bandwidth between single- and multi-GPU systems. For the latter case, the efficiency factor $f_{PCIe,multi}$ may be expressed as

$$f_{PCIe,multi}(n) = \begin{cases} 0.227 \log(n) - 1.235 & (n \leq 16K) \\ 1 & (n > 16K) \end{cases}. \quad (15)$$

$T_{net}$ is the time for transferring data of j-particles through gigabit network. In the ring communication topology, $T_{net}$ may be expressed as

$$T_{net}(N) = t_{net} \cdot (N_{gpu} - 1) \cdot n, \quad (16)$$

where $t_{net}$ is the time for transferring a single particle. It may be expressed as $t_{net}$ = 28/$BW_{net}$, where $BW_{net}$ is the average measured bandwidth of gigabit network. The estimated value of $t_{PCIe,multi}$ and $t_{net}$ are recorded in Table 2. Note that for the dual-GPU case, since two GPUs are installed in the same node, we set $T_{net}$ = 0 in our performance model.

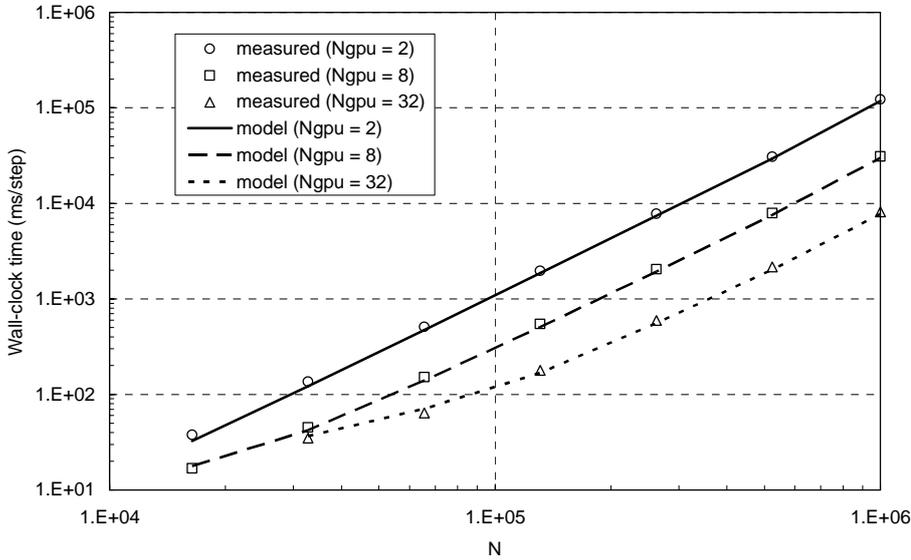

Fig. 19. The wall-clock time per step as a function of the total number of particles (N) in the multi-GPU system. Solid, dashed, and dotted curves are the results predicted by the performance model for $N_{gpu}$ = 2, 8, and 32, respectively. Open circles, squares, and triangles are the measured results for $N_{gpu}$ = 2, 8, and 32, respectively.

Fig. 19 shows the wall-clock time per step predicted by Eq. (14) for $N_{gpu}$ = 2, 8, 32, along with the measured performance for comparison. Again, the agreement between modeled and measured performance is very good. It indeed captures the feature of lower slope when $N/N_{gpu}$ < 4K (see Section 4.3).

Fig. 20 shows the relative ratios of $T_{host}$, $T_{PCIe, multi}$, $T_{net}$, and $T_{GPU}$ in multi-GPU performance model with $N_{gpu}$ = 2, 8, 32. It can be seen that although the total calculation time is generally dominated by $T_{GPU}$, the time spent on network communication is non-negligible in some cases. $T_{net}/T_{multi}$ exceeds 20% for N $\leqq$ 128K in 32-GPU simulations and 10% for N $\leqq$ 64K in 8-GPU simulations. Also note that although $T_{PCIe,\,multi}$ plays a minor role in performance modeling, $T_{PCIe,\,multi}/T_{multi}$ still exceeds 9% for N $\leqq$ 128K in 32-GPU simulations and 9% for N $\leqq$ 32K in 8-GPU simulations. Finally, it shows that $T_{host}$ is negligible in all cases.

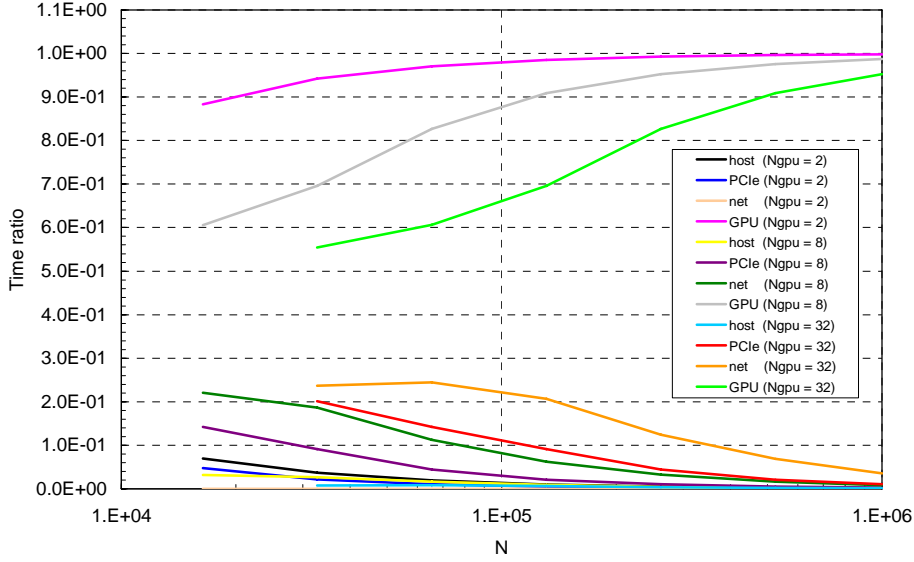

Fig. 20. The relative ratios of $T_{host}$, $T_{PCIe,\,multi}$, $T_{net}$, and $T_{GPU}$ as a function of the total number of particles (N) predicted by the performance model of multi-GPU system. The cases of $N_{gpu}$ = 2, 8, and 32 are plotted, respectively.

## 6. Discussion

In this section, we address on the comparison of performance between GraCCA and GRAPE system, the stability of GraCCA, and finally a discussion of future work.

6.1. Comparison with GRAPE

6.1.1. Single-GPU system

The GRAPE-6A board with four GRAPE-6 processor chips has become available since 2005 (Fukushige et al., 2005). It is a single PCI card (currently a GRAPE PCI-X board is available) attached to a host computer. The theoretical peak performance of a single GRAPE-6A board is 131.3 GFLOPS for acceleration and jerk calculation. The maximum number of particles it can handle is up to 256k. A single GRAPE-6A card costs about $6K.

In comparison with GRAPE-6A, the NVIDIA GeForce 8800 GTX graphic card transfers data between PC and GPU memory through PCI-Express x16 slot, which has a theoretical bandwidth of 4 GB/s in each direction. It is about 30 times higher than PCI bus. This rather high bandwidth eliminates the bottleneck in data communication between host computer and computing coprocessor. On the other hand, the measured performance of GRAPE-6A card is currently limited by its communication speed. Therefore, by implementing the gravitational N-body calculation in a single 8800 GTX graphic card, we have achieved a computing power of about 250 GFLOPS for N $\geqq$ 16K. Moreover, with 768MB on-board memory it can store up to 10M particles, at a cost of only about $630 as of late 2006. In other words, we have achieved about 18 times better performance-per-dollar than GRAPE-6A board.

6.1.2. Multi-GPU system

The GRAPE-6 system was built in 2002 (Makino et al., 2003). It comprises 64 processor boards, each of which has a theoretical peak speed of about 1 TFLOPS. Thus, the GRAPE-6 has a total performance of about 64 TFLOPS and can handle up to 32 million particles. However, in practical situation, the peak performance is marginally above 20 TFLOPS, which is limited by the bandwidth of the communication between different host computers. Each processor board costs about $50K. So, regardless of the cost of other components, the 64 processor boards in GRAPE-6 still cost about $3.2M.

There are several existing PC-GRAPE clusters using GRAPE-6A. For examples, Fukushige et al. (2005) have built a 24-node cluster in the University of Tokyo. The University of Amsterdam has constructed a 4-node cluster named MODEST[2]. Harfst et al. (2007) have reported the performance of two 32-node clusters, which are located in the Rochester Institute of Technology (RIT) and in the Astronomisches Rechen-institut (ARI) at the University of Heidelberg, respectively. Here we only compare the performance of GraCCA with the PC-GRAPE cluster at RIT. It consists of 32 nodes and each of which is attached with a GRAPE-6A PCI card. The theoretical performance of this system is roughly 4 TFLOPS, and the total cost is about $450K. For simulating a large number of particles, it achieves a measured performance of about 3 TFLOPS.

In comparison with GRAPE-6 and GRAPE-6A cluster in RIT, our GPU cluster consisting of 32 GeForce 8800 GTX graphic cards has a measured performance of about 7.1 TFLOPS, which is more than two times higher than that of GRAPE-6A cluster in RIT. Although it is still about one-third of the measured performance of GRAPE-6, our system only costs about $32K (including all components within the

cluster) and can store up to 320M particles. Stated in another way, we have achieved a performance-per-dollar about 35.5 times better than that of GRAPE-6 system and 33.3 times better than that of GRAPE-6A cluster in RIT. Furthermore, in contrast to GRAPE-6 and GRAPE-6A which are special-purpose computers, modern graphic cards are fully programmable. So our GPU cluster is more flexible and can actually serve as a general-purpose computer. Finally, the modern graphic cards only support single-precision accuracy at present (NVIDIA, 2007) (the NVIDIA Corporation has announced that GPUs supporting double-precision accuracy will become available in late 2007). By contrast, the GRAPE hardware uses a 64-bit fixed-point format to accumulate the acceleration (Makino et al., 2003), and therefore results in a higher accuracy than GPU. This issue has been addressed by Belleman et al. (2007), Hamada & Iitaka (2007), and Portegies Zwart et al. (2007).

6.2. Stability of GraCCA

Although commercial graphic cards are generally thought to have a relatively short time between failures, we have not experienced such instability. For example, the core collapse simulation for 64K particles took about 1 month, and the run had not experienced any system crash. It was paused several times due only to manual interruptions. However, improper coding in GPU program may easily and instantly lead to system idle or system crash. On the contrary, improper coding in CPU program generally only results in a forcible process termination.

6.3. Future outlook

Currently, we only use the shared time-step scheme for the purpose of performance measurements. This scheme is expected to be inaccurate for orbits of close pairs and may have an artifact of collision. In order to maximize the efficiency of direct N-body simulation as well as to improve accuracy for close pairs, we will adopt the individual time-step scheme along with block time-step algorithm[7]. Two issues may arise when we switch to this scheme. First, as illustrated in Fig. 11, the performance of single GPU drops dramatically for N < 4K in our current implementation, which is mainly caused by the insufficient number of threads. Although this is not a problem in shard time-step scheme since we are more interested in large-N systems, it can suppress the performance in individual time-step scheme, where the number of i-particles to be updated in each step is much smaller than optimal number of particles (N). One solution to this problem is to equally divide the force calculation of a single i-particle into several parts, and each part is computed by one thread. In this way, we can keep the number of threads in GPU large enough even for small N (Belleman et al., 2007). The second issue is that the ratio of communication time in network ($T_{net}$) to total calculation time ($T_{multi}$) becomes worse. There are two ways to get around this problem. One is to use a faster network, such as Infiniband or Myrinet. Another is to adopt a more efficient scheme for parallel force computation (Makino, 2002; Harfst et al., 2007).

---

[7] The scheme of parallel individual time steps along with block time-step algorithm has been implemented after the submission of this paper and the results and comparison will be reported in a separate paper.

In addition, instead of using the direct-summation scheme for the entire gravitational system, we may treat the direct N-body computation as a GPU kernel to be embedded in the general cosmology computation. Most cosmology problems deal with dark matter particles, which are inherently collisionless where the gravitational force is given by the mean field. However, at the densest core regions, the computational of the mean field is limited by grid resolution. The dynamical range of spatial resolution is therefore severely limited to at most 4 orders of magnitude in mean field calculations. Conventionally, one circumvents this resolution problem by employing the direct N-body computation only for those particles in the densest cores to increase the local force resolution. The direct N-body part turns out to be the most time consuming. This is where the GPU computation comes to play. The GPU program can replace the existing CPU sub-routine of direct N-body calculations. The replacement will shorten the cosmology computation time by a sizable factor.

Finally, Gualandris et al. (2007) have presented the first highly parallel, grid-based N-body simulations. It opens a paradigm for the GraCCA system to connect with the grid-computing community in the future.

## Acknowledgments

We would like to thank Hui-Min Yao for helping building the GraCCA system. We are grateful to Zhe Fan for stimulating discussions on GPGPU and Simon Portegies Zwart for providing valuable suggestions on this paper. We also thank the GIGABYTE Corporation for providing partial support for the hardware in the GraCCA system. This work is supported in part by the National Science Council of Taiwan under the grant NSC96-2112-M-002-002.